# AN ELECTRIC CIRCUIT LAB

# FOR GENERAL PHYSICS CLASSES

Kyle Rogers

Luciano Fleischfresser

Oklahoma School of Science and Mathematics

Autry Technology Center

1201 West Willow Road

Enid, OK 73703

1/15/2007 1:15:01 PM

**Overview**

This paper describes an electric circuit lab exercise that offers various ways of comparing the equivalent electrical resistance of three-resistor circuits. It requires the use of two multi-meters, a power supply, cables and connectors, an assortment of ceramic resistors, and breadboards. The lab lets students verify Ohm's law through graphical analysis, and perform comparisons with calculated counterparts. The lab can be shortened as well as be made more complex. Suggestions are presented on how to simplify the assignment, and on how to make it a more challenging task for students.

**Theory**

We used three ceramic resistors with gold band tolerance (5 %), and color-coded[1] brown-black-brown [$R_1 = (100 \pm 5)\Omega$], red-red-brown [$R_2 = (220 \pm 11)\Omega$], and yellow-violet-brown [$R_3 = (470 \pm 23.5)\Omega$] to construct a series configuration and two series-parallel configurations shown in figures 1, 2 and 3.

The equivalent resistance expressions for the three given configurations are as follows:

$$R_{eq1} = R_1 + R_2 + R_3 \tag{1}$$

$$R_{eq2} = R_1 + \frac{R_2 \cdot R_3}{R_2 + R_3} \tag{2}$$

$$R_{eq3} = \frac{(R_1 + R_2) \cdot R_3}{R_1 + R_2 + R_3} \tag{3}$$

We treat the individual resistors as unknowns in (1), (2), and (3). We then measure the equivalent resistances $R_{eq1}$, $R_{eq2}$ and $R_{eq3}$ when set on the breadboard (table I). Solving the system of three equations with three unknowns for $R_3$, a quadratic equation is obtained as follows:

$$R_3^2 - R_{eq1} \cdot R_3 + R_{eq1} \cdot R_{eq3} = 0 \tag{4}$$

To have real solutions for $R_3$ in (4), $R_{eq1} \geq 4 \cdot R_{eq3}$. This condition imposes a limit on the choice of resistor sets. After obtaining $R_3$ from (4), one can go back to (1) and (2) to calculate $R_1$ and $R_2$. Table II shows the results along with the true resistances of each resistor used.

**Experimental Procedure**

Figure 4 shows the set-up for measuring voltage and current on all configurations. We close the circuit and obtain five pairs of data points by varying the voltage from 3 to 7 volts across it. The results for all configurations are shown in figure 5. It serves to verify the ohmic nature of the ceramic resistors. Finding the slope on the plots provides a way to double-check the electrical resistance measurements presented earlier.

We follow the recommendations of [2] to determine error bounds on the voltage and current data. It is assumed that uncertainties due to scale resolution and rated accuracy are the dominant ones. The multi-meter used, a METEX® M-3800, has rated accuracies of $\pm 0.5\%$ for the 20 V range used for all set-ups, and also for the 20 mA range used for the series configuration. It has a rated accuracy of $\pm 1.2\%$ for the 200 mA range used for the mixed arrangements (configurations #2 and #3). Table III presents the data collected with the combined uncertainties for each measurement.

**Discussion**

This assignment is intended after the class has spent some time learning and applying the concepts involved. Therefore, students should already have some familiarity with the concepts of electricity. There are plenty of resources available if one is searching for materials to tie with this activity[3, 4, 5, 6, 7, 8, 9, 10, 11, 12, 13, 14, 15, and 16].

The lab lets pupils build simple resistor circuits, practice electrical measurements, graphical analysis, and algebraic derivations. Instructors have options to simplify or increase the complexity of the lab depending on their needs and intended outcomes. For example, one could use four instead of three resistors and introduce other arrangements. A more challenging activity would be to measure the electrical *current* through one resistor in a given configuration, and compare it with the calculated value from Ohm's law and, perhaps Kirchoff's rules, depending on the complexity of the circuit analyzed. If a simpler lab is the goal, one might skip the graphical analysis, and have students compare the measured, calculated, and nominal resistances. We asked students to show the derivation of equation (4) to get extra credit in the lab.

The idea for this lab outgrew from the circuit lab used at OSSM. Kyle Rogers suggested that students should discover the individual resistances, instead of have them given from the outset. In the original lab, they take the three *known* resistors and are asked to make all possible configurations that have the $100\,\Omega$ resistor (13 in all). They are then asked to make equivalent resistance and voltage measurements to compare with the calculated counterparts.

Generally, our students enjoy working with circuits. We think the lab described here helps them if used after exposure to the concepts and practice with electrical measurements. As an added bonus, it provides a worthy mathematical challenge.

**List of References**

**Short Biographies:**

**Kyle Rogers** is a senior at Enid High School. He graduated from OSSM-ATC in 2006. He plans to study biomedical engineering at the University of Oklahoma.

**Luciano Fleischfresser** has been teaching general physics and mechanics since 2000 at OSSM-ATC. He received his Ph.D. from the University of Oklahoma.

**Physics Lab., OSSM – ATC, Enid, OK 73703;**

**l_fle@alumni.ou.edu**

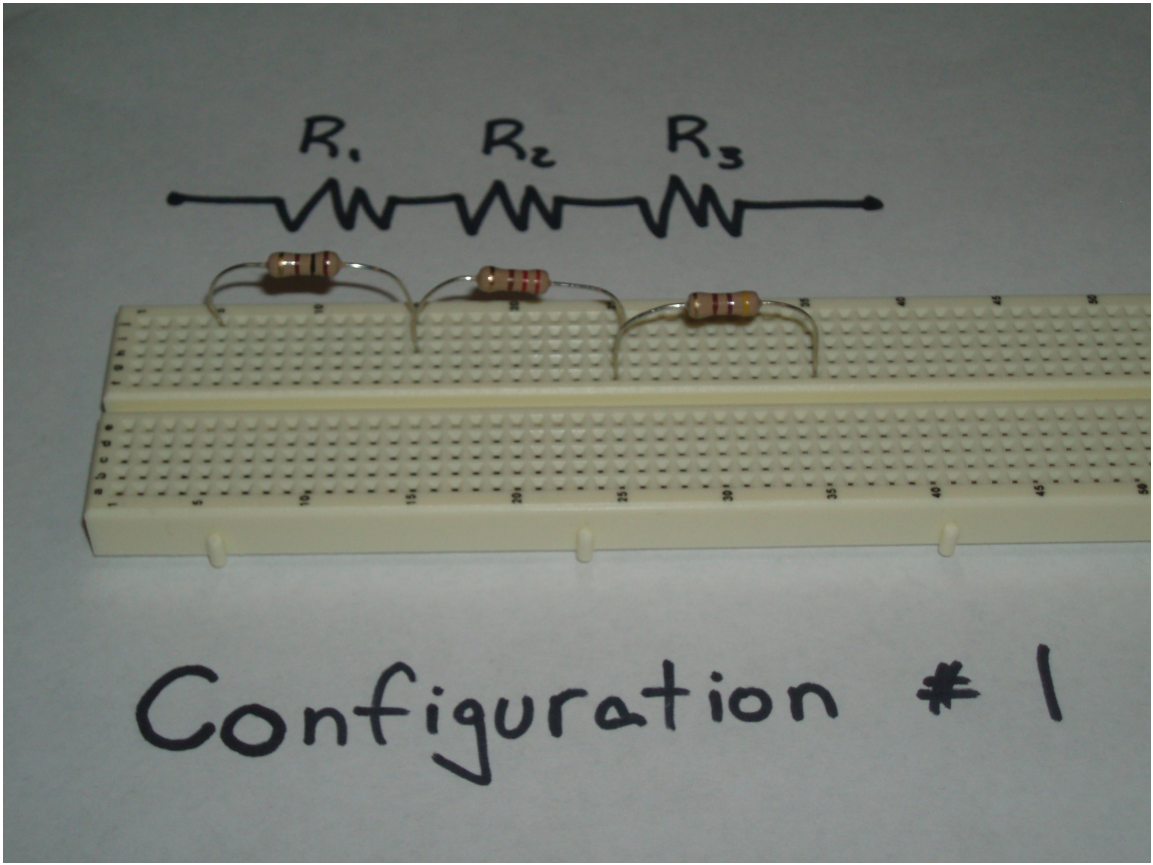

**Figure 1: The three resistors in series on the breadboard (configuration # 1).**

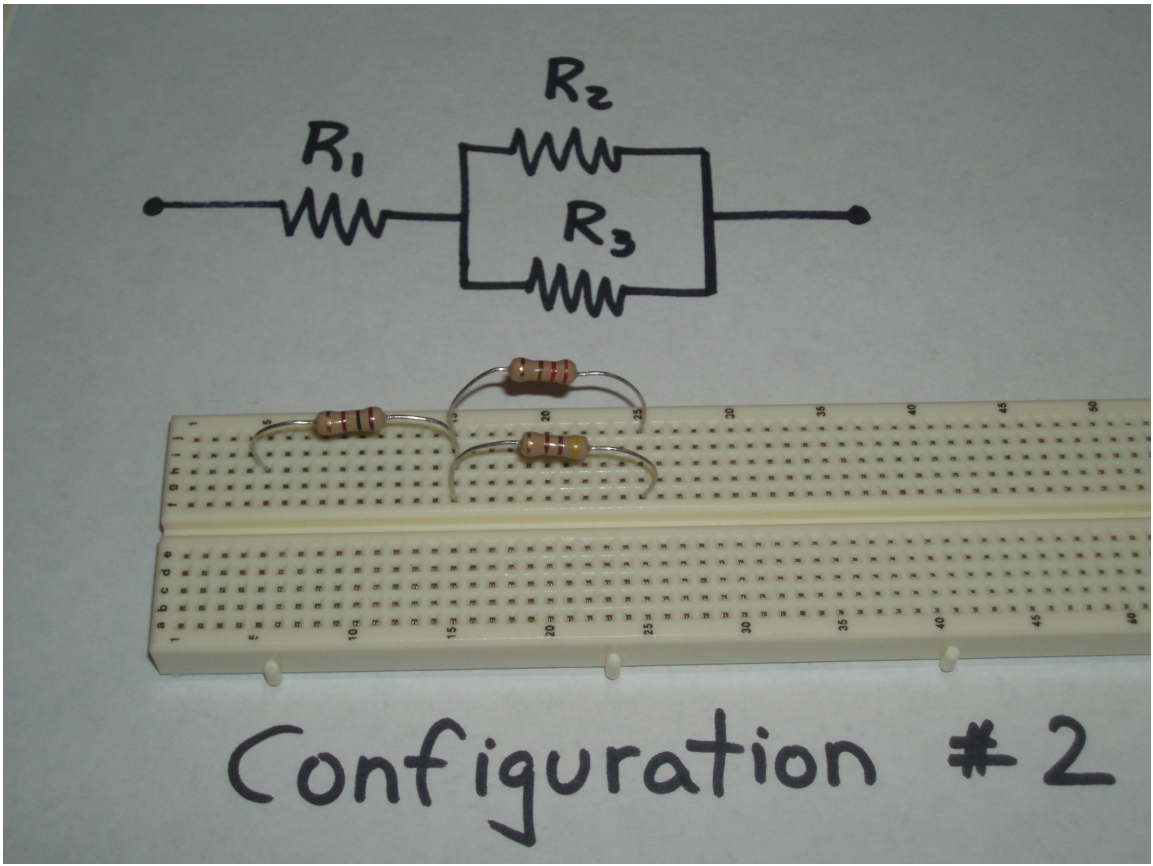

**Figure 2: The first series-parallel configuration on the breadboard (configuration #2).**

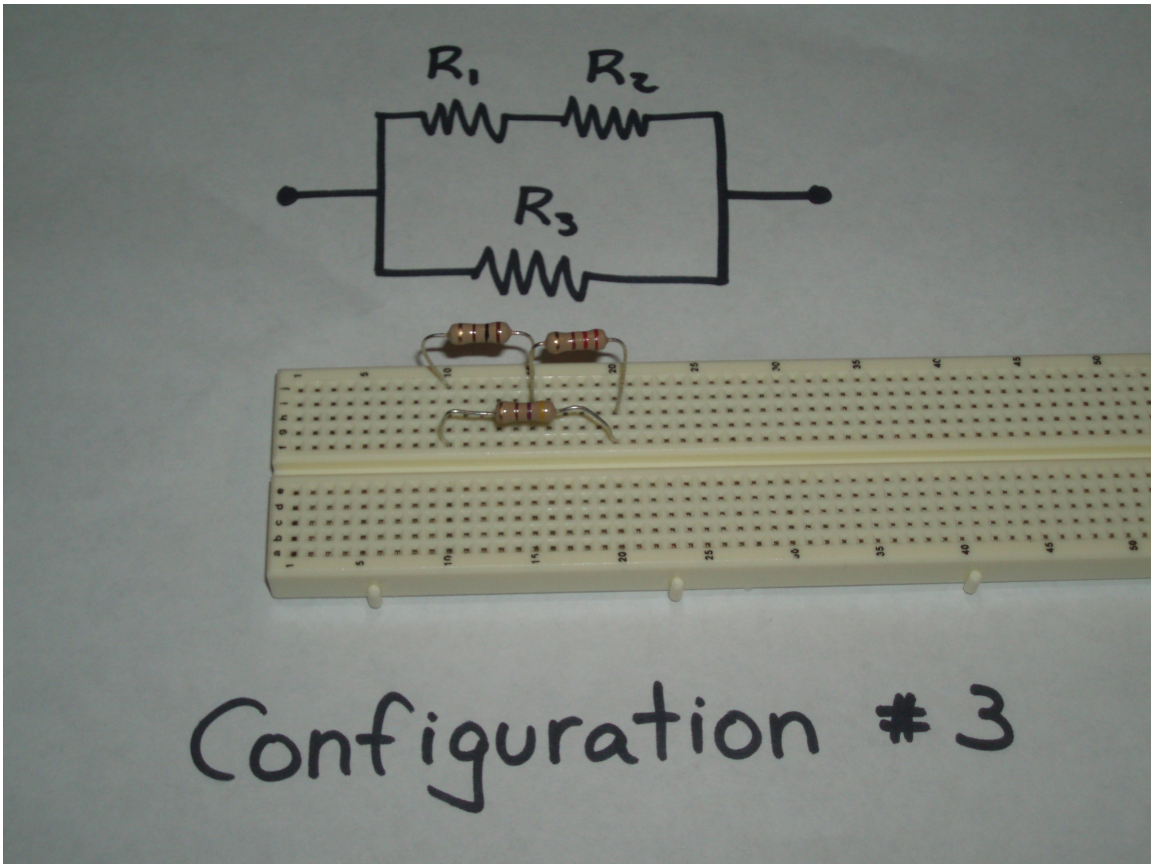

**Figure 3: The second series-parallel configuration on the breadboard (configuration # 3).**

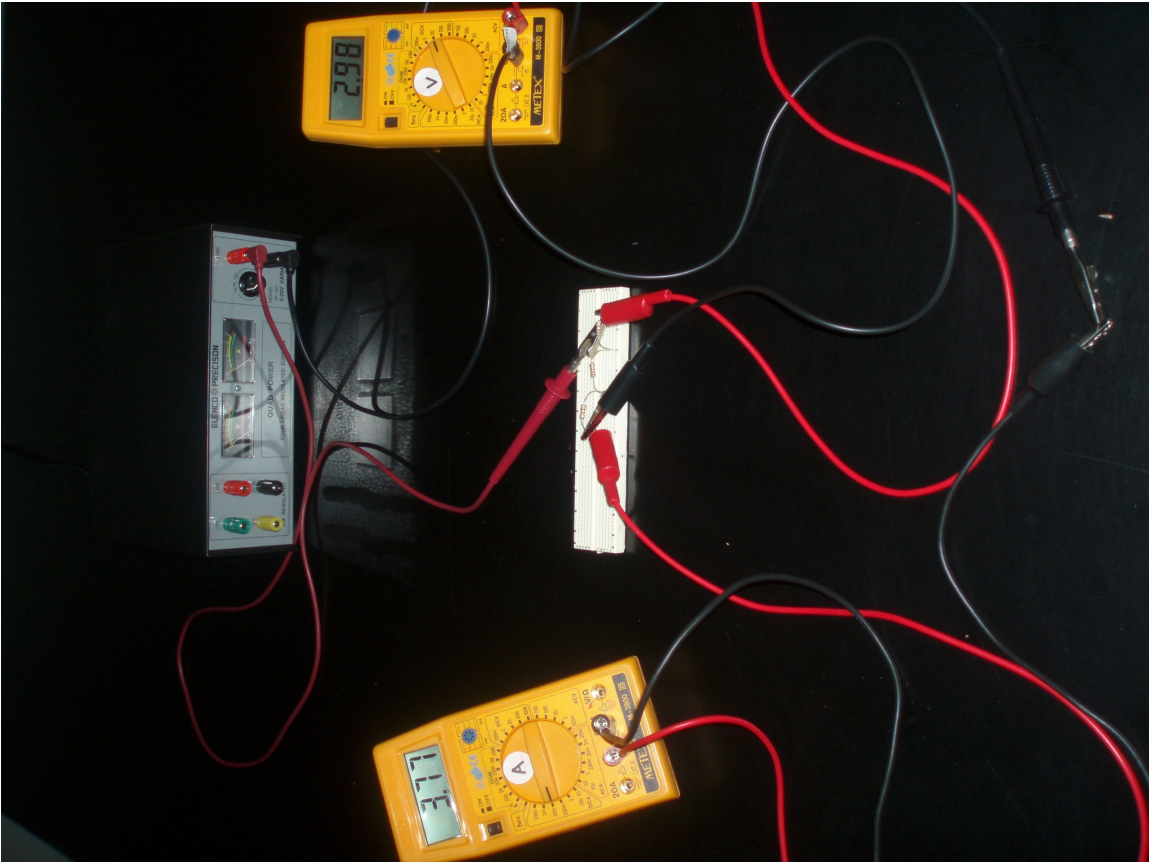

**Figure 4: Set-up for voltage and current measurements.**

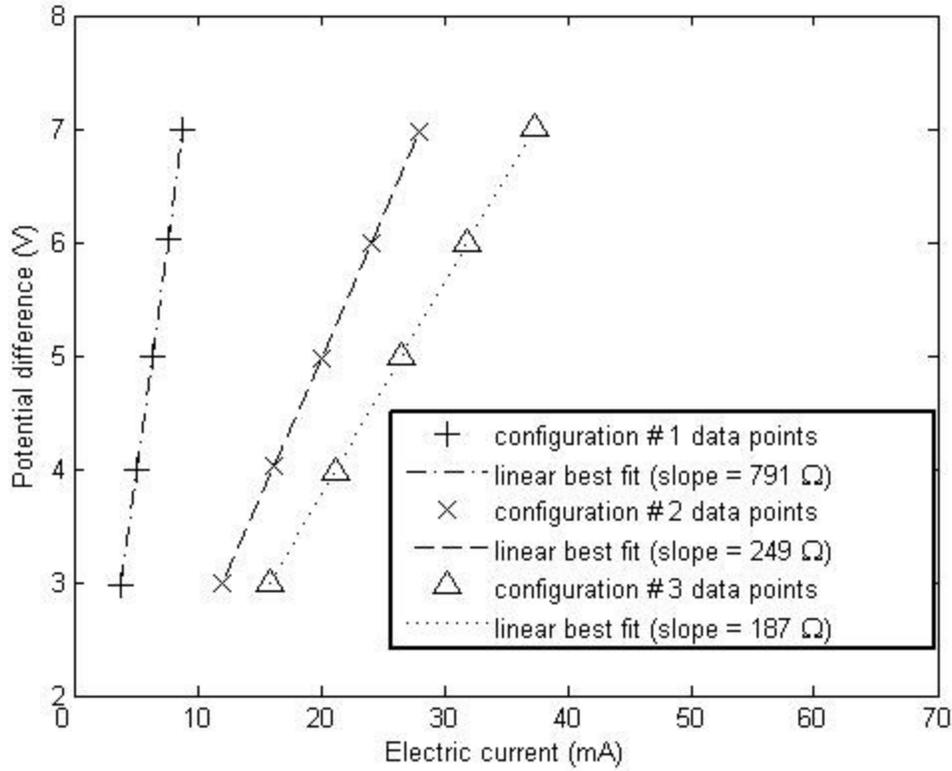

**Figure 5: Electrical current versus voltage for the three configurations. Also shown is the linear fit for each with the slope given.**

**Table I:** *Measured* **equivalent resistances for configurations #1, #2 and #3.**

| Configuration #1 | Configuration #2 | Configuration #3 |
|---|---|---|
| 789 Ω | 248 Ω | 188 Ω |

**Table II:** *Calculated* **and** *true* **values for each ceramic resistor.**

| Calculated | True (conf. #1) | True (conf. #2) | True (conf. #3) |
|---|---|---|---|
| 105 Ω | 101 Ω | 102 Ω | 101 Ω |
| 204 Ω | 216 Ω | 215 Ω | 214 Ω |
| 480 Ω | 470 Ω | 463 Ω | 464 Ω |

Table III: Voltage and current measurements for all circuits. Also shown is the combined uncertainty due to rated accuracy and scale resolution for each measurement.

| Configuration #1 | | Configuration #2 | | Configuration #3 | |
|---|---|---|---|---|---|
| (V) | (mA) | (V) | (mA) | (V) | (mA) |
| 2.980 ± 0.015 | 3.760 ± 0.019 | 2.990 ± 0.015 | 12.00 ± 0.15 | 2.990 ± 0.015 | 15.90 ± 0.19 |
| 4.000 ± 0.020 | 5.040 ± 0.025 | 4.030 ± 0.020 | 16.20 ± 0.20 | 3.980 ± 0.020 | 21.20 ± 0.26 |
| 4.990 ± 0.025 | 6.300 ± 0.032 | 4.980 ± 0.025 | 20.00 ± 0.24 | 5.000 ± 0.025 | 26.60 ± 0.32 |
| 6.020 ± 0.030 | 7.600 ± 0.038 | 5.990 ± 0.030 | 24.10 ± 0.29 | 5.990 ± 0.030 | 31.90 ± 0.38 |
| 6.990 ± 0.035 | 8.830 ± 0.044 | 6.970 ± 0.035 | 28.00 ± 0.34 | 7.010 ± 0.035 | 37.40 ± 0.45 |